# Integrated Guidance and Control for Lunar Landing using a Stabilized Seeker

Brian Gaudet*
*University of Arizona, 1127 E. Roger Way, Tucson Arizona, 85721*

Roberto Furfaro[†]
*University of Arizona, 1127 E. Roger Way, Tucson Arizona, 85721*

We develop an integrated guidance and control system that in conjunction with a stabilized seeker and landing site detection software can achieve precise and safe planetary landing. The seeker tracks the designated landing site by adjusting seeker elevation and azimuth angles to center the designated landing site in the sensor field of view. The seeker angles, closing speed, and range to the designated landing site are used to formulate a velocity field that is used by the guidance and control system to achieve a safe landing at the designated landing site. The guidance and control system maps this velocity field, attitude, and rotational velocity directly to a commanded thrust vector for the lander's four engines. The guidance and control system is implemented as a policy optimized using reinforcement meta learning. We demonstrate that the guidance and control system is compatible with multiple diverts during the powered descent phase, and is robust to seeker lag, actuator lag and degradation, and center of mass variation induced by fuel consumption. We outline several concepts of operations, including an approach using a preplaced landing beacon.

## I. Introduction

$S$AFE planetary landing requires the ability to identify and track a suitable landing site in real time during the powered descent phase. The selected landing site should be at a low slope with respect to the planetary equipotential surface, free of hazards, and also be located such that it satisfies mission objectives. The Apollo Lunar missions allowed the lander pilot to steer towards a manually selected landing site by manipulating a control stick until the designated landing site (DLS) appeared in a window fixed reticle [1], but the actual trajectory flown to reach that point in a manner consistent with a soft landing was automated by the guidance and control system. The Chinese Chang'e 3 mission [2] used coarse hazard detection from greyscale images during the powered descent, and once the lander reached an altitude of 100 m, hovered while a hazard free landing site was selected, after which the lander diverted horizontally until directly above the selected landing site, and then continued straight down to the surface. Finally, the Morpheus project [3] demonstrated integration of flash LIDAR based hazard detection with guidance and control on Earth using a hazard field with a mix of terrain hazards and safe landing sites. During the powered descent the landing site detection system will likely change the DLS as distance to surface decreases and resolution increases. Consequently, a suitable guidance and control system should be capable of multiple divert maneuvers during the powered descent phase.

In this work we develop a guidance, navigation, and control (GN&C) architecture that uses measurements from seeker hardware to create a reference velocity field. The guidance and control (G&C) system then tracks this velocity field to achieve a soft landing at the DLS, mapping velocity field tracking error, change in attitude, and rotational velocity directly to commanded thrust for the lander's four thrusters. The system architecture is illustrated in Fig. 1, where the G&C system is implemented as a policy optimized using meta reinforcement learning (meta-RL). In the meta-RL framework, an agent instantiating the policy learns through episodic simulated experience over an ensemble of environments covering the expected distribution of mission scenarios, sensor lag, actuator degradation, variation in system time constants, and other factors. The policy is implemented as a deep neural network parameterized by $\theta$ that maps observations to actions $\mathbf{u} = \pi_\theta(\mathbf{o})$, and is optimized using a customized version of proximal policy optimization (PPO) [4]. Adaptation is achieved by including a recurrent network layer [5] with hidden state $\mathbf{h}$ in both the policy and value function networks. Maximizing the PPO objective function requires learning hidden layer parameters $\boldsymbol{\theta}_h$ that result in $\mathbf{h}$ evolving in response to the history of $\mathbf{o}$ and $\mathbf{u}$ in a manner that facilitates fast adaptation to an environment sampled

---

*Engineer, Department of Systems and Industrial Engineering, E-mail:briangaudet@mac.com

[†]Professor, Department of Systems and Industrial Engineering, Department of Aerospace and Mechanical Engineering

from the ensemble, and generalization to novel environments. The deployed policy will then adapt to off-nominal flight conditions. Importantly, the network parameters remain fixed during deployment with adaptation occurring through the evolution of **h**. Although it can take several days to optimize a meta-RL policy, the deployed policy can be run forward by computing several small matrix multiplications, which should take less than 10 ms on a modern flight computer. Meta-RL has been recently applied to aerospace GN&C applications including planetary powered descent [6, 7], exo-atmospheric intercept [8], and asteroid close proximity operations [9, 10].

Referring to Fig. 1, this work does not develop the landing site selection software (LSSS), and we assume a LSSS implementation that can identify and track a safe landing site during the powered descent phase, potentially shifting the DLS as range decreases and measurements become more accurate. The initial attitude is the attitude at the start of the powered descent phase with respect to the Lunar equipotential surface in the vicinity of the landing region. Note that it is possible to formulate a guidance system that does not require attitude with respect to the surface, but just the change in attitude during the powered descent phase. However, it is likely that the LSSS will require an estimate of the lander's attitude with respect to the surface in order to estimate slope, so we gave the G&C system access to the lander's pitch and roll with respect to the Lunar surface in addition to the change in attitude.

Figure 2 illustrates the stabilized seeker. The seeker is mounted to a mechanically stabilized platform that is connected to the lander body using three gimbals, allowing the platform attitude to be independent of the lander attitude. These platform gimbals are adjusted by the stabilization gimbal controller in order to keep the seeker platform at a constant attitude during the powered descent phase, thus creating an inertial reference frame. Without stabilization, lander rotation would appear as DLS motion, complicating the task of the G&C system. The sensor is in turn mounted to the seeker platform using elevation and azimuth gimbals that are adjusted by the pointing gimbal controller in a manner that keeps the DLS centered in the sensor field of view (FOV). Note that when the DLS is centered in the sensor FOV, the sensor boresight axis will be pointing at the DLS. Although the arrangement might seem complex, it is mature technology that has been used for decades in missile guidance systems [11].

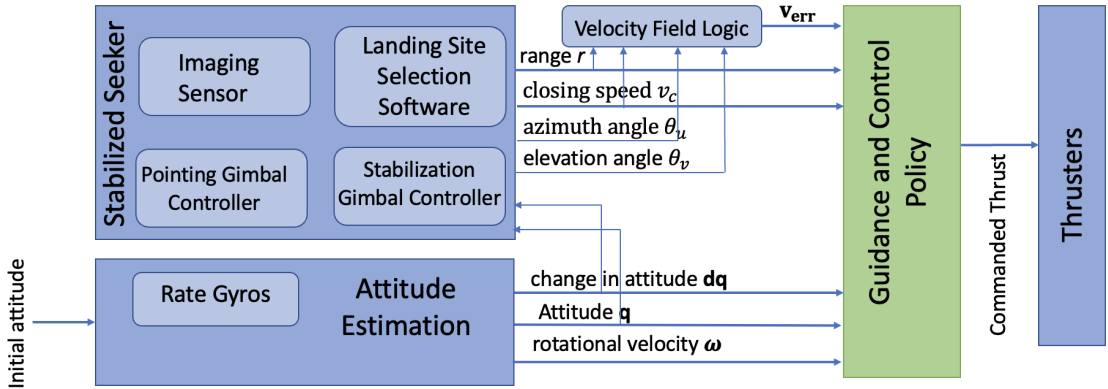

**Fig. 1 System Diagram**

Although in this work we use a mechanically stabilized and pointed seeker, it is also possible to computationally stabilize the seeker platform using the estimated change in attitude during the powered descent phase to rotate the elevation and azimuth angles back to where they would be if measured at the lander's attitude at the start of the powered descent phase. This technique is mechanically simpler, as the sensor elevation and azimuth gimbals are mounted directly to the lander body. Finally, it is also possible to use a strapdown seeker architecture, where the sensor itself is fixed in the lander body frame. A detailed implementation of a strapdown seeker for an exo-atmospheric intercept application can be found in [8].

For the GN&C architecture considered in this work, the imaging sensor must be capable of measuring the range and closing speed to the location on the planetary surface that is intersected by the sensor boresight axis. For some implementations this might require multiple sensors pointing in the same direction. For example, if the LSSS uses camera images, then a Doppler LIDAR rangefinder can be coupled with a camera, and in the steady state where the seeker is locked to the DLS, the rangefinder can measure range and closing speed to the DLS. Similarly, using a Doppler flash LIDAR sensor, the center LIDAR element's range and closing speed can be averaged to provide range and closing speed to the DLS. In the following (see Section IV.B), we show how the closing speed, elevation, and azimuth angles can be used to formulate a velocity field, and that this velocity field in conjunction with the estimated change in lander attitude and rotational velocity can be mapped directly to commanded thrust levels by the integrated G&C policy.

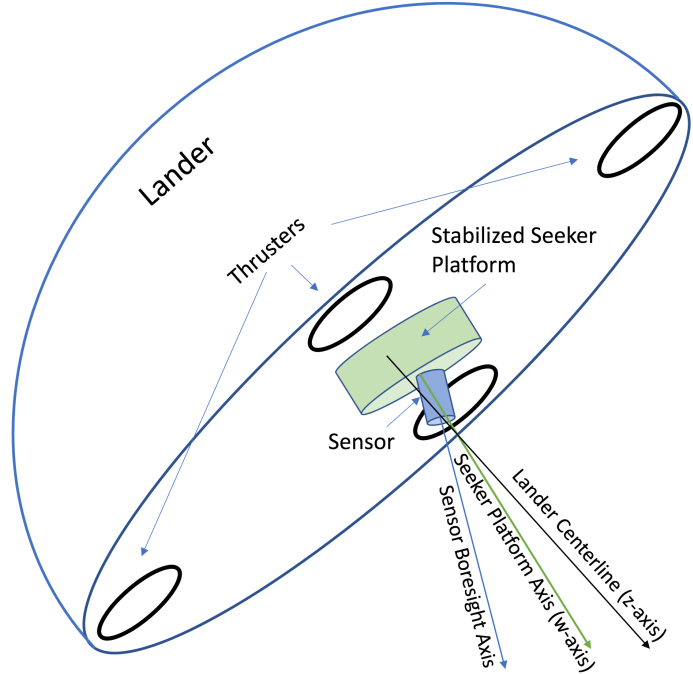

**Fig. 2  Gimbaled Seeker Implementation**

A particularly simple concept of operations that is appealing for a lander carrying human passengers, high value cargo, or both, is to have a rover pre-place an optical beacon at a safe landing site. The lander's seeker would employ a sensor consisting of a camera and Doppler LIDAR rangefinder, with the optical axis of each sensor aligned so that when the optical beacon is centered in the camera field of view, the rangefinder gives range to the beacon. This simplifies the LSSS, which now only needs to differentiate between the beacon and the lunar surface. Mission risk is also reduced, as it is very unlikely that the LSSS will mistake a natural surface feature for the landing beacon. In contrast, there is a higher probability of error for the case where the LSSS must identify and track a safe landing site by imaging the Lunar surface without the aid of a beacon. An alternative concept that would work for a two part landing system (command module and lander) would be for the orbiting command module to use a laser to tag the desired landing site from orbit as the lander descends. In this case the seeker would lock to the light reflected back from the landing site. On Earth this approach is used for laser guided missiles, and is a mature technology.

To our knowledge, this is the first published using missile seeker hardware as part of a GN&C system for the powered descent phase of a planetary landing, although it has been applied to asteroid close proximity operations in [9]. The remainder of the paper is organized as follows. Section II formulates the powered descent phase problem, giving initial conditions, equations of motion, divert scenarios, and describing the seeker and thruster models. Section III then gives an overview of the reinforcement learning framework. Section IV describes the meta-RL formulation of the guidance and landing policies and describes multiple experiments used to validate the integrated G&C system.

## II. Problem Formulation

### A. Powered Descent Phase

The goal is to achieve a safe landing somewhere within a region of interest on the Lunar surface. A de-orbit and braking burn puts the lander on an open loop trajectory towards this landing region. Prior to the start of the powered descent phase, the LSSS identifies a safe landing site, which we choose as the origin of the coordinate system that will be used for simulation purposes to specify the position of the lander $\mathbf{r}_L$, and $\mathbf{r}_T$, the position of the current DLS. As the lander approaches the current DLS, the LSSS will be able to better assess the suitability of potential landing sites, and may determine that the current DLS is no longer safe. This will require a divert maneuver to a newly selected DLS. In this work we assume that four diverts will be required at ranges of 1500m, 1000m, 500m, and 100m from the current DLS, with each divert randomly generated as shown in Eqs. (1a) and (1b). Let $\mathbf{r}_{LT} = \mathbf{r}_L - \mathbf{r}_T$. Then the downrange and crossrange elements of the divert vector are each a maximum of 10% of $\|\mathbf{r}_{LT}\|$ (the current distance between the lander and DLS), and the altitude divert is a maximum of 5% of the current distance between the lander and DLS. $\mathcal{U}(x, y, d)$ is a uniformly distributed $d$ dimensional random variable bounded by $x$ and $y$.

$$\Delta_{\text{divert}} = \|\mathbf{r}_{LT}\| \begin{bmatrix} 0.1 & 0.1 & 0.05 \end{bmatrix} \mathcal{U}(-1, 1, 3) \tag{1a}$$

$$\mathbf{r}_{T_{\text{new}}} = \mathbf{r}_{T_{\text{old}}} + \Delta_{\text{divert}} \tag{1b}$$

The simulated initial conditions at the start of the powered descent phase are as shown in Table 1. Defining an optimal heading as the lander's velocity vector $\mathbf{v}_L$ colinear with $-\mathbf{r}_{LT}$, the initial lander velocity will have a random heading error $\theta_v$ between 0 and 10 degrees. The lander's nominal initial attitude is such that the -Z body-frame axis is aligned with the line of sight to target. This nominal attitude is perturbed at the start of each episode such that the angle between the -Z body frame axis and line of sight to target varies uniformly between 0.0 degrees and 10 degrees. To avoid confusion, note that the angles in Table 1 have no preferred direction, i.e., they are oriented randomly in $\mathbb{R}^3$.

**Table 1   Initial Conditions**

| Parameter | min | max |
|---|---|---|
| Downrange $\mathbf{r}_L[0]$ (m) | 1500 | 2000 |
| Crossrange $\mathbf{r}_L[1]$ (m) | -500 | 500 |
| Altitude $\mathbf{r}_L[2]$ (m) | 2000 | 2200 |
| Speed $V_{\text{init}}$ (m/s) | 40 | 50 |
| Heading Error $\theta_v$ (degrees) | 0 | 10 |
| Attitude Error $\theta_q$ (degrees) | 0 | 10 |
| Mass $m$ (kg) | 1900 | 2000 |
| Diagonal Inertia Tensor Perturbation $\Delta\mathbf{J}_{\text{diag}}$ (kg-m$^2$) | -10 | 10 |
| Off-Diagonal Inertia Tensor Perturbation $\Delta\mathbf{J}_{\text{off}}$ (kg-m$^2$) | -1 | 1 |
| Seeker time constant $\tau_{\text{seeker}}$ (s) | 0.2 | 0.2 |
| Engine time constant $\tau_{\text{ctrl}}$ (s) | 0.2 | 0.2 |

### B. Equations of Motion

For purposes of modeling the lander's moments of inertia, we model the lander as a uniform density ellipsoid, with inertia matrix given by

$$\mathbf{J} = \frac{m}{5} \begin{bmatrix} b^2 + c^2 & 0 & 0 \\ 0 & a^2 + c^2 & 0 \\ 0 & 0 & a^2 + b^2 \end{bmatrix} \tag{2}$$

where $a = 2$ m, $b = 2$ m, and $c = 1$ m correspond to the body frame $x$, $y$, and $z$ axes. $m$ is the lander's mass, which is updated as shown in Eq. (8c). The diagonal elements of $\mathbf{J}$ are perturbed at the start of each episode by adding a uniformly distributed random variable $\mathcal{U}(-\Delta\mathbf{J}_{\text{diag}}, \Delta\mathbf{J}_{\text{diag}}, 3)$ and the off-diagonal elements are similarly perturbed but in a way that keeps $\mathbf{J}$ positive semi-definite.

At the start of the powered descent phase, the lander has a mass ranging from 1900 to 2000 kg and four throttleable thrusters with a minimum and maximum thrust magnitude of 500 N and 2500 N respectively. The four thrusters are located in the lander body frame as shown in Table 2, where $x$, $y$, and $z$ are the body frame axes. Roll is about the $x$-axis, yaw is about the $z$-axis, and pitch is about the $y$-axis. Note that this thruster configuration does not allow any direct control of the rotational velocity around the $z$-axis. However, the lander's yaw will change during the trajectory, but due to coupling with pitch via roll rather than due to torque caused by thrust.

The force $\mathbf{F}_B$ and torque $\mathbf{L}_B$ in the lander's body frame for a given commanded thrust depends on the placement of the thrusters in the lander structure. We can describe the placement of each thruster through a body-frame direction vector $\mathbf{d}$ and position vector $\mathbf{r}$, both in $\mathbb{R}^3$. The direction vector is a unit vector giving the direction of the body frame force that results when the thruster is fired. The position vector gives the body frame location with respect to the lander centroid, where the force resulting from the thruster firing is applied for purposes of computing torque, and in general the center of mass $\mathbf{r}_{com}$ varies with time as fuel is consumed. For a lander with $k$ thrusters, the body frame force and torque associated with one or more thrusters firing is then as shown in Equations (3a) through (3c), where $\mathbf{u}^{(i)}$ is the commanded thrust for thruster $i$, $\mathbf{d}^{(i)}$ the direction vector for thruster $i$, $\mathbf{r}^{(i)}$ the position of thruster $i$, and $\mathbf{F}_B^{(i)}$ the force contribution for thruster $i$. The total body frame force and torque are calculated by summing the individual forces and torques.

**Table 2    Body Frame Thruster Locations.**

| Thruster | x (m) | y (m) | z (m) |
|----------|-------|-------|-------|
| 1 | 0 | -2 | -1 |
| 2 | 0 | 2 | -1 |
| 3 | -2 | 0 | -1 |
| 4 | 2 | 0 | -1 |

$$\mathbf{F}_B^{(i)} = \mathbf{d}^{(i)} \mathbf{u}^{(i)} \tag{3a}$$

$$\mathbf{F}_B = \sum_{i=1}^{k} \tilde{\mathbf{F}}_B^{(i)} \tag{3b}$$

$$\mathbf{L}_B = \sum_{i=1}^{k} (\mathbf{r}^{(i)} - \mathbf{r}_{com}) \times \tilde{\mathbf{F}}_B^{(i)} \tag{3c}$$

The instantaneous center of mass is modeled as shown in Equation 4, where $\mathbf{r}_{com}(t)$ is the center of mass at time $t$. Here $\boldsymbol{\zeta}$ is a uniformly distributed unit direction vector generated at the start of each episode, $\alpha$ is a scaling factor, $f_{used}$ is the fuel used from the start of the powered descent phase up to time $t$, and $f_{max} = 200$kg is the amount of fuel remaining in the lander at the start of the powered descent phase. Note that this is not meant to accurately model the shift in center of mass due to fuel consumption, but instead to demonstrate that the G&C system can tolerate dynamic shifts in center of mass during the powered descent phase. We use $\alpha = 0.1$ which corresponds to a 0.1 m maximum possible shift in the lander's center of mass, or 5% of the lander's major axis.

$$\mathbf{r}_{com}(t) = \alpha \boldsymbol{\zeta} \frac{f_{used}(t)}{f_{max}} \tag{4}$$

The dynamics model uses the lander's current attitude $\mathbf{q}$ to convert the body frame thrust vector to the inertial frame as shown in Equation (5) where $\mathbf{C}_{BN}(\mathbf{q})$ is the direction cosine matrix mapping the inertial frame to body frame obtained from the current attitude parameter $\mathbf{q}$.

$$\mathbf{F}_N = [\mathbf{C}_{BN}(\mathbf{q})]^T \mathbf{F}_B \tag{5}$$

The rotational velocities $\boldsymbol{\omega}$ are then obtained by integrating the Euler rotational equations of motion, as shown in Equation (6), where $\mathbf{L}_B$ is the body frame torque as given in Equation (3b), and $\mathbf{J}$ is the lander's inertia tensor. Note we have included a term that models a rotation induced by a changing inertia tensor, which in general is time varying as the lander consumes fuel. Specifically, the inertia tensor is recalculated at each time step to account for fuel consumption, but we do not modify the inertia tensor to account for changes in the lander's center of mass.

$$\mathbf{J}\dot{\boldsymbol{\omega}} = -\tilde{\boldsymbol{\omega}}\mathbf{J}\boldsymbol{\omega} - \dot{\mathbf{J}}\boldsymbol{\omega} + \mathbf{L}_B \tag{6}$$

The lander's attitude is then updated by integrating the differential kinematic equations shown in Equation (7), where the lander's attitude is parameterized using the quaternion representation and $\omega_i$ denotes the $i^{th}$ component of the rotational velocity vector $\boldsymbol{\omega}$.

$$\begin{bmatrix} \dot{q_0} \\ \dot{q_1} \\ \dot{q_2} \\ \dot{q_3} \end{bmatrix} = \frac{1}{2} \begin{bmatrix} q_0 & -q_1 & -q_2 & -q_3 \\ q_1 & q_0 & -q_3 & q_2 \\ q_2 & q_3 & q_0 & -q_1 \\ q_3 & -q_2 & q_1 & q_0 \end{bmatrix} \begin{bmatrix} 0 \\ \omega_0 \\ \omega_1 \\ \omega_2 \end{bmatrix} \tag{7}$$

The translational motion is modeled as shown in Eqs. (8a) through (8c), where $\mathbf{F}^N$ is the inertial frame force as given in Eq. (5), $g_{\text{ref}} = 9.8$ m/s$^2$, $\mathbf{g} = \begin{bmatrix} 0 & 0 & -1.63 \end{bmatrix}$ m/s$^2$ is used for the Earth's Moon, $I_{\text{sp}} = 225$ s, and the spacecraft's mass is $m$.

$$\dot{\mathbf{r}}_L = \mathbf{v}_L \tag{8a}$$

$$\dot{\mathbf{v}}_L = \frac{\mathbf{F}^N}{m} + \mathbf{g} \tag{8b}$$

$$\dot{m} = -\frac{\sum_i^k \|\mathbf{F}^{B^{(i)}}\|}{I_{\text{sp}} g_{\text{ref}}} \tag{8c}$$

The navigation system provides updates to the guidance system every 0.2 s, and we integrate the equations of motion using fourth order Runge-Kutta integration with a time step of 0.05 s.

## C. Engine Model

The output of the G&C policy $\mathbf{u}_\pi = \pi(\mathbf{o}) \in \mathbb{R}^4$ determines the thrust of the four lander engines according to Eqs. (9a) and (9b). $\mathbf{u}_\pi$ is multiplied by the maximum allowed thrust $u_{\text{max}}$ and then clipped so that $\mathbf{u}$ falls between the minimum and maximum allowed thrust. At the start of each episode, partial actuator failure is modeled by randomly selecting one of the four engines using the index $i_{\text{AF}}$ and randomly determining if the episode results in actuator failure. If actuator failure occurs, $fail$ is set to True. The selected engine's thrust is then scaled by $s_{\text{AF}}$

$$\mathbf{u}_{\text{cmd}}^{(i)} = \text{clip}(u_{\text{max}} \, \mathbf{u}_\pi, u_{\text{min}}, u_{\text{max}}) \tag{9a}$$

$$\mathbf{u}_{\text{AF}}^{(i_{\text{AF}})} = \begin{cases} \mathbf{u}_{\text{cmd}}^i \, s_{\text{AF}} & \text{if } fail \text{ and } i == i_{\text{AF}} \\ \mathbf{u}_{\text{cmd}}^{(i)} & \text{otherwise} \end{cases} \tag{9b}$$

$$\tag{9c}$$

Actuator delay is then modeled by integrating Eq. (10a), where $\tau_{\text{ctrl}}$, is the actuator time constant, and $\mathbf{u}$ is the control inputs used in Eq. (3a).

$$\dot{\mathbf{u}} = \frac{\mathbf{u}_{\text{AF}} - \mathbf{u}}{\tau_{\text{ctrl}}} \tag{10a}$$

## D. Seeker Model

We assume that the lander is equipped with a stabilized seeker with a 90 degree field of regard, mounted on the bottom of the lander. Note that in this work we do not model the pointing or stabilization controllers (see Fig. 1), but assume that the seeker perfectly tracks the DLS with some gimbal lag $\tau_{\text{seeker}}$. Thus, neglecting the gimbal lag, the elevation and azimuth angles are calculated assuming the seeker boresight axis is pointing at the DLS. Prior to the powered descent phase, the seeker elevation angle $\theta_u$ and azimuth angle $\theta_v$ are both set to zero, and the seeker platform's attitude is adjusted such that the DLS is centered in the sensor FOV. Therefore, at the start of the powered descent phase, the seeker gimbal angles are zero and the seeker platform is at an attitude where the DLS is centered in the sensor's field of view. During the powered descent phase, the attitude of the seeker platform remains constant unless reset, and the seeker angles $\theta_u$ and $\theta_v$ vary as the pointing controller keeps the sensor boresight aligned with the DLS.

To simulate the observed seeker angles $\theta_u$ and $\theta_v$ as the seeker tracks the target from the stabilized reference frame, we can define the angles between the seeker boresight axis and the seeker reference frame $u$ and $v$ axes as the seeker angles $\theta_u$ and $\theta_v$. Let $\mathbf{r}_{\text{TL}} = \mathbf{r}_T - \mathbf{r}_L$ denote the relative position of the DLS with respect to the lander. Further, let the superscript N denote an inertial reference frame fixed to the Lunar surface and the superscript S denote the seeker platform reference frame. $\mathbf{C}_{\text{SN}}(\mathbf{q}_0)$ is then the direction cosine matrix (DCM) mapping from N to S, with $\mathbf{q}_0$ being the lander's attitude when the seeker is reset, e.g., at the start of the powered descent phase. We can now transform the DLS relative position from the inertial reference frame $\mathbf{r}_{\text{TL}}^N$ into the seeker reference frame as shown in Eq. (11).

$$\mathbf{r}_{\text{TL}}^S = [\mathbf{C}_{\text{SN}}(\mathbf{q}_0)] \mathbf{r}_{\text{TL}}^N \tag{11}$$

Defining the line of sight unit vector in the seeker reference frame as $\hat{\lambda}^S = \frac{\mathbf{r}_{TL}^S}{\|\mathbf{r}_{TL}\|^S}$ and the seeker frame unit vectors $u = \begin{bmatrix} 1 & 0 & 0 \end{bmatrix}$, $v = \begin{bmatrix} 0 & 1 & 0 \end{bmatrix}$, and $w = \begin{bmatrix} 0 & 0 & 1 \end{bmatrix}$ we can then compute the observed un-lagged seeker angles as the orthogonal projection of the seeker frame line of sight (LOS) vector onto $u$ and $v$ as shown in Eqs. (12a) and (12b).

$$\tilde{\theta}_u = \arcsin(\hat{\lambda}^S \cdot \hat{u}) \tag{12a}$$

$$\tilde{\theta}_v = \arcsin(\hat{\lambda}^S \cdot \hat{v}) \tag{12b}$$

We also assume that the sensor can provide range $r$ and closing speed $v_c$. These un-lagged observations are computed as shown in Eqs. (13a) and (13b).

$$\tilde{r} = \|\mathbf{r}_{TL}\| \tag{13a}$$

$$\tilde{v}_c = -\frac{\mathbf{r}_{TL} \cdot \mathbf{v}_{TL}}{\|\mathbf{r}_{TL}\|} \tag{13b}$$

The seeker gimbal actuation is then modeled using a first order lag by integrating Eqs.(14a) through (14d)

$$\dot{\theta}_u = \frac{\tilde{\theta}_u - \theta_u}{\tau_{\text{seeker}}} \tag{14a}$$

$$\dot{\theta}_v = \frac{\tilde{\theta}_v - \theta_v}{\tau_{\text{seeker}}} \tag{14b}$$

$$\dot{r} = \frac{\tilde{r} - r}{\tau_{\text{seeker}}} \tag{14c}$$

$$\dot{v}_c = \frac{\tilde{v}_c - v_c}{\tau_{\text{seeker}}} \tag{14d}$$

## III. Background: Reinforcement Learning Framework

In the reinforcement learning framework, an agent learns through episodic interaction with an environment how to successfully complete a task using a policy that maps observations to actions. The environment initializes an episode by randomly generating a ground truth state, mapping this state to an observation, and passing the observation to the agent. The agent uses this observation to generate an action that is sent to the environment; the environment then uses the action and the current ground truth state to generate the next state and a scalar reward signal. The reward and the observation corresponding to the next state are then passed to the agent. The process repeats until the environment terminates the episode, with the termination signaled to the agent via a done signal. Trajectories collected over a set of episodes (referred to as rollouts) are collected during interaction between the agent and environment, and used to update the policy and value functions. The interface between agent and environment is depicted in Fig. 3.

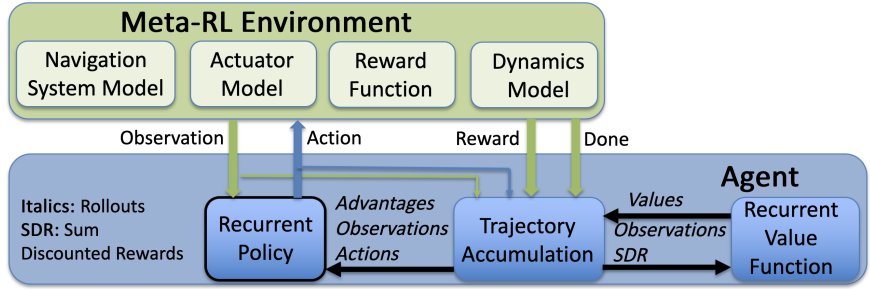

**Fig. 3  Environment-Agent Interface**

A Markov Decision Process (MDP) is an abstraction of the environment, which in a continuous state and action space, can be represented by a state space $\mathcal{S}$, an action space $\mathcal{A}$, a state transition distribution $\mathcal{P}(\mathbf{x}_{t+1}|\mathbf{x}_t, \mathbf{u}_t)$, and a reward function $r = \mathcal{R}(\mathbf{x}_t, \mathbf{u}_t)$, where $\mathbf{x} \in \mathcal{S}$ and $\mathbf{u} \in \mathcal{A}$, and $r$ is a scalar reward signal. We can also define a partially observable MDP (POMDP), where the state $\mathbf{x}$ becomes a hidden state, generating an observation $\mathbf{o}$ using an observation function $O(\mathbf{x})$ that maps states to observations. The POMDP formulation is useful when the observation consists of sensor outputs. In the following, we will refer to both fully observable and partially observable environments as POMDPs, as an MDP can be considered a POMDP with an identity function mapping states to observations.

Meta-RL differs from generic reinforcement learning in that the agent learns over an ensemble of POMPDs. These POMDPs can include different environmental dynamics, aerodynamic coefficients, actuator failure scenarios, mass and inertia tensor variation, and varying amounts of sensor distortion. Optimization within the meta-RL framework results in an agent that can quickly adapt to novel POMDPs, often with just a few steps of interaction with the environment. There are multiple approaches to implementing meta-RL. In [12], the authors design the objective function to explicitly make the model parameters transfer well to new tasks. In [13], the authors demonstrate state of the art performance using temporal convolutions with soft attention. And in [14], the authors use a hierarchy of policies to achieve meta-RL. In this work, we use an approach similar to [15] using a recurrent policy and value function. Note that it is possible to train over a wide range of POMDPs using a non-meta RL algorithm. Although such an approach typically results in a robust policy, the policy cannot adapt in real time to novel environments. In this work, we implement meta-RL using proximal policy optimization (PPO) [4] with both the policy and value function implementing recurrent layers in their networks. After training, although the recurrent policy's network weights are frozen, the hidden state will continue to evolve in response to a sequence of observations and actions, thus making the policy adaptive. In contrast, a policy without a recurrent layer has behavior that is fixed by the network parameters at test time.

The PPO algorithm used in this work is a policy gradient algorithm which has demonstrated state-of-the-art performance for many reinforcement learning benchmark problems. PPO approximates the Trust Region Policy Optimization method [16] by accounting for the policy adjustment constraint with a clipped objective function. The objective function used with PPO can be expressed in terms of the probability ratio $p_k(\boldsymbol{\theta})$ given by,

$$p_k(\boldsymbol{\theta}) = \frac{\pi_{\boldsymbol{\theta}}(\mathbf{u}_k|\mathbf{o}_k)}{\pi_{\boldsymbol{\theta}_{\text{old}}}(\mathbf{u}_k|\mathbf{o}_k)} \tag{15}$$

The PPO objective function is shown in Equations (16a) through (16c). The general idea is to create two surrogate objectives, the first being the probability ratio $p_k(\boldsymbol{\theta})$ multiplied by the advantages $A_{\mathbf{w}}^{\pi}(\mathbf{o}_k, \mathbf{u}_k)$ (see Eq. (17)), and the second a clipped (using clipping parameter $\epsilon$) version of $p_k(\boldsymbol{\theta})$ multiplied by $A_{\mathbf{w}}^{\pi}(\mathbf{o}_k, \mathbf{u}_k)$. The objective to be maximized $J(\boldsymbol{\theta})$ is then the expectation under the trajectories induced by the policy of the lesser of these two surrogate objectives.

$$\text{obj1} = p_k(\boldsymbol{\theta})A_{\mathbf{w}}^{\pi}(\mathbf{o}_k, \mathbf{u}_k) \tag{16a}$$

$$\text{obj2} = \text{clip}(p_k(\boldsymbol{\theta})A_{\mathbf{w}}^{\pi}(\mathbf{o}_k, \mathbf{u}_k), 1 - \epsilon, 1 + \epsilon) \tag{16b}$$

$$J(\boldsymbol{\theta}) = \mathbb{E}_{p(\tau)}[\min(\text{obj1}, \text{obj2})] \tag{16c}$$

This clipped objective function has been shown to maintain a bounded Kullback-Leibler (KL) divergence [17] with respect to the policy distributions between updates, which aids convergence by ensuring that the policy does not change drastically between updates. Our implementation of PPO uses an approximation to the advantage function that is the difference between the empirical return and a state value function baseline, as shown in Equation 17, where $\gamma$ is a discount rate and $r$ the reward function, described in Section **??**.

$$A_{\mathbf{w}}^{\pi}(\mathbf{x}_k, \mathbf{u}_k) = \left[\sum_{\ell=k}^{T} \gamma^{\ell-k} r(\mathbf{o}_{\ell}, \mathbf{u}_{\ell})\right] - V_{\mathbf{w}}^{\pi}(\mathbf{x}_k) \tag{17}$$

Here the value function $V_{\mathbf{w}}^{\pi}$ is learned using the cost function given by

$$L(\mathbf{w}) = \frac{1}{2M} \sum_{i=1}^{M} \left(V_{\mathbf{w}}^{\pi}(\mathbf{o}_k^i) - \left[\sum_{\ell=k}^{T} \gamma^{\ell-k} r(\mathbf{u}_{\ell}^i, \mathbf{o}_{\ell}^i)\right]\right)^2 \tag{18}$$

In practice, policy gradient algorithms update the policy using a batch of trajectories (roll-outs) collected by interaction with the environment. Each trajectory is associated with a single episode, with a sample from a trajectory collected at step $k$ consisting of observation $\mathbf{o}_k$, action $\mathbf{u}_k$, and reward $r_k(\mathbf{o}_k, \mathbf{u}_k)$. Finally, gradient ascent is performed on $\boldsymbol{\theta}$ and gradient descent on $\mathbf{w}$ and update equations are given by

$$\mathbf{w}^+ = \mathbf{w}^- - \beta_{\mathbf{w}} \nabla_{\mathbf{w}} L(\mathbf{w})|_{\mathbf{w}=\mathbf{w}^-} \tag{19}$$

$$\boldsymbol{\theta}^+ = \boldsymbol{\theta}^- + \beta_{\boldsymbol{\theta}} \nabla_{\boldsymbol{\theta}} J(\boldsymbol{\theta})|_{\boldsymbol{\theta}=\boldsymbol{\theta}^-} \tag{20}$$

where $\beta_{\mathbf{w}}$ and $\beta_{\boldsymbol{\theta}}$ are the learning rates for the value function, $V_{\mathbf{w}}^{\pi}(\mathbf{o}_k)$, and policy, $\pi_{\boldsymbol{\theta}}(\mathbf{u}_k|\mathbf{o}_k)$, respectively.

In our implementation of PPO, we adaptively scale the observations and servo both $\epsilon$ and the learning rate to target a KL divergence of 0.001.

# IV. Methods

## A. Two Segment Powered Descent Phase

We obtained the best performance by using a two segment powered descent phase, where the powered descent phase is split into a guidance segment and a landing segment. The guidance segment starts at the beginning of the powered descent phase and ends when the lander reaches an altitude 5 m above the altitude of the DLS, at which point the landing segment begins. Each segment uses a separately optimized policy for the G&C system. The policy used in the guidance segment attempts to track a velocity field that keeps the lander's line of sight to the DLS aligned with its velocity vector. This suffices to bring the lander within 5 m of the DLS at a speed of less than 2 m/s. Note that the lander does not end the guidance segment directly above the DLS, because once the lander's relative altitude falls below 5 m, it will still be slightly downrange of the target. The landing segment policy then attempts to bring the lander straight down towards the Lunar surface such that contact occurs at a level attitude (pitch and roll both less than 10 degrees) with negligible rotational velocity (all elements of $\boldsymbol{\omega}$ less than 10 degrees / s) and a velocity vector that is close to perpendicular with the Lunar equipotential surface at the DLS, as quantified by a glideslope of at least 80 degrees.

## B. Guidance Segment meta-RL Formulation

Given a stabilized seeker, hitting the DLS is straightforward, as we could use proportional navigation [18], and reward the agent for minimizing the rate of change of the seeker angles; this proved effective in an exo-atmospheric intercept application [8]. However, in the planetary landing application, we want to achieve both a soft and accurate landing. Parallel navigation [19] can be modified by formulating and tracking a velocity field that both keeps the lander's velocity vector colinear with the line of sight to the DLS and tracks a reference speed that is a function of range to the DLS. This approach was used successfully in [6], where the guidance system had access to the ground truth lander position and velocity. However, constructing a velocity field from seeker outputs requires estimating the lander inertial frame velocity vector, e.g., using an unscented Kalman filter [20].

In this work we took a different approach and developed a method that allows direct mapping of the seeker angles, range, and closing speed to a reference velocity field, that when tracked, results in both an accurate and soft landing. Our approach periodically (every two seconds) resets the seeker platform such that the platform w-axis is aligned with the DLS, while the gimbaled seeker tracks the target continuously at the navigation update frequency (5 Hz). Note that this reset would be accomplished by adjusting the seeker platform attitude until the DLS is centered in the sensor FOV, as described in Section II.D, but without resetting the gimbal angles $\theta_u$ and $\theta_v$. To see why this periodic reset is necessary, consider the start of the powered descent phase where the lander is on a velocity vector with a 10° heading error. When the seeker platform is reset at the start of the powered descent phase (as described in Section II.D) the seeker platform (with $\theta_u = \theta_v = 0$) w-axis will be pointing directly at the desired landing site. However, as the heading error is corrected during the trajectory, this seeker platform attitude is no longer optimal, and as the lander gets within a hundred meters of the landing site, the landing site is no longer within the sensor FOV, even with the seeker gimbal angles at maximum. This problem is compounded when multiple diverts occur during the powered descent phase. With these periodic resets of the seeker platform, minimizing the seeker angles $\theta_u$ and $\theta_v$ keeps $\mathbf{r}_{TL}$ colinear with $\mathbf{v}_{TL}$. Note that these periodic resets involve only a small change in seeker platform attitude, except for the case of divert maneuvers, as can be seen in Section V, Fig. 8.

The reference velocity field is formulated using the range $r = \|\mathbf{r}_{TL}\|$, closing speed $v_c$ and the seeker angles $\theta_u$ and $\theta_v$, as shown in Eqs. (21a) through (21d), where we use $\tau_{vref} = 25$s, and $v_{c_o}$ is the lander's closing speed $v_c$ at the start of the powered descent phase. To be clear, $\mathbf{v}_{\lambda}$ is *not* the lander velocity vector in the seeker reference frame, but is the reconstructed line of sight unit vector in the seeker reference frame multiplied by the closing speed.

$$t_{\text{go}} = \frac{r}{v_c} \tag{21a}$$

$$\mathbf{v}_\lambda = v_c \left[ \sin\theta_u \quad \sin\theta_v \quad \sqrt{1 - \sin^2\theta_u - \sin^2\theta_v} \right] \tag{21b}$$

$$\mathbf{v}_{\text{ref}} = v_{c_0} \left( 1 - \exp\left( -\frac{t_{\text{go}}}{\tau_{\text{vref}}} \right) \right) \begin{bmatrix} 0 & 0 & 1 \end{bmatrix} \tag{21c}$$

$$\mathbf{v}_{\text{err}} = \mathbf{v}_\lambda - \mathbf{v}_{\text{ref}} \tag{21d}$$

The velocity field tracking error $\mathbf{v}_{\text{err}}$ is then used to compute shaping rewards at each step of interaction between the agent and environment. These shaping rewards encourage the agent to track the velocity field. The full reward function used to optimize the guidance segment policy is given in Eqs. (22a) through (22e), where $r_{\text{ctrl}}$ penalizes the agent for excess control effort (recall that $k$ is the number of thrusters) and $r_{\text{term}}$ is given at the end of the episode if the relative altitude with respect to the DLS is less than 5 m (which terminates the episode), the lander is within 10 m of the DLS, and the lander's speed is less than 2 m/s. Here $\mathbf{r}_{\text{LT}} = -\mathbf{r}_{\text{TL}}$. We denote the Euler 321 attitude parameterization as $\mathbf{q}_{321}$, where $\mathbf{q}_{321}[1]$ corresponds to pitch and $\mathbf{q}_{321}[2]$ corresponds to roll. An episode is terminated if the absolute value of the pitch or roll exceeds 85°, and the agent is given a penalty of -100. In this work we used $\alpha = -0.5$, $\beta = -0.01$, $\eta = 0.01$, and $\kappa = 20$.

$$r_{\text{shaping}} = \eta + \alpha \|\mathbf{v}_{\text{err}}\| \tag{22a}$$

$$r_{\text{ctrl}} = \beta \frac{\mathbf{F}_B}{k \, u_{\text{max}}} \tag{22b}$$

$$r_{\text{term}} = \begin{cases} \kappa, & \text{if } \|\mathbf{r}_{\text{LT}}[2]\| < 5 \text{ m and } \|\mathbf{r}_{\text{LT}}\| < 10 \text{ m and } \|\mathbf{v}_{\text{LT}}\| < 2 \text{ m/s} \\ 0, & \text{otherwise} \end{cases} \tag{22c}$$

$$r_{\text{pen}} = \begin{cases} -100, & \text{if } \mathbf{q}_{321}[1] > 85° \text{ or } \mathbf{q}_{321}[2] > 85° \\ 0, & \text{otherwise} \end{cases} \tag{22d}$$

$$r = r_{\text{shaping}} + r_{\text{ctrl}} + r_{\text{pen}} + r_{\text{term}} \tag{22e}$$

The agent observation is shown in Eq. 23, where $\mathbf{dq}$ is computed by integrating the quaternion kinematic equations given in Eq. (24), with $\mathbf{dq}_0 = [1, 0, 0, 0]$. qsub is the quaternion subtraction operation, and $\mathbf{q}^S$ is the attitude of the stabilized seeker platform. Note that since we periodically reset the seeker platform attitude, the seeker platform only provides an inertial reference frame between resets. Consequently, we need to augment the observation with $\text{qsub}(\mathbf{q}, \mathbf{q}^S)$ so that the agent is aware of changes to the attitude of the stabilized seeker platform.

$$\mathbf{o} = \begin{bmatrix} \mathbf{v}_{\text{err}} & t_{\text{go}} & \|\mathbf{r}_{\text{TL}}\| & \mathbf{dq} & \text{qsub}(\mathbf{q}, \mathbf{q}^S) & \mathbf{q}_{321}[1] & \mathbf{q}_{321}[2] & \omega \end{bmatrix} \tag{23}$$

$$\begin{bmatrix} \dot{dq}_0 \\ \dot{dq}_1 \\ \dot{dq}_2 \\ \dot{dq}_3 \end{bmatrix} = \frac{1}{2} \begin{bmatrix} dq_0 & -dq_1 & -dq_2 & -dq_3 \\ dq_1 & dq_0 & -dq_3 & dq_2 \\ dq_2 & dq_3 & dq_0 & -dq_1 \\ dq_3 & -dq_2 & dq_1 & dq_0 \end{bmatrix} \begin{bmatrix} 0 \\ \omega_0 \\ \omega_1 \\ \omega_2 \end{bmatrix} \tag{24}$$

During optimization, rollouts are collected over 60 episodes of interaction and environment and used to update the policy and value function.

## C. Landing Segment meta-RL Formulation

For the landing segment policy optimization, an episode terminates once the lander's altitude falls below that of the DLS. The landing segment policy is optimized using only a control effort penalty and two terminal rewards, as shown in Eqs. (25a) through (25d), where the "any" operator is True only if all elements of the vector satisfy the inequality, whereas $r_{\text{term1}}$ rewards the agent for executing a landing that maximizes glideslope and minimizes pitch, roll, rotational velocity, and landing speed. $r_{\text{term1}}$ rewards the agent for meeting the criterion for a safe landing. An episode is terminated if the absolute value of the pitch or roll exceeds $85°$, and the agent is given a penalty of -100.

$$r_{\text{ctrl}} = \beta \frac{\mathbf{F}_B}{k \; u_{\max}} \tag{25a}$$

$$r_{\text{term1}} = \begin{cases} \kappa \exp\left(-\frac{\left\|\begin{bmatrix} \frac{5 \, \|\mathbf{v}_{\text{LT}}[0:1]\|}{\mathbf{v}_{\text{LT}}[2]} & \mathbf{v}_L & \mathbf{q}_{321}[1] & \mathbf{q}_{321}[2] & \omega \end{bmatrix}\right\|^2}{\sigma_L^2}\right), & \text{if } \mathbf{r}_{\text{LT}}[2] < 5 \text{ m} \\ 0, & \text{otherwise} \end{cases} \tag{25b}$$

$$r_{\text{term2}} = \begin{cases} \kappa, & \text{if } \mathbf{r}_{\text{LT}}[2] < 5 \text{ m and } \|\mathbf{v}_{\text{LT}}\| < 2 \text{ m/s and } |\mathbf{q}_{321}[2]| < 10° \text{ and} \\ & |\mathbf{q}_{321}[1]| < 10° \text{ and all}(\omega < 10°/\text{s}) \text{ and } \arctan\left(\frac{\mathbf{v}_{\text{LT}}[2]}{\|\mathbf{v}_{\text{LT}}[0:1]\|}\right) > 80°) \\ 0, & \text{otherwise} \end{cases} \tag{25c}$$

$$r = r_{\text{ctrl}} + r_{\text{term1}} + r_{\text{term2}} \tag{25d}$$

Because the landing policy does not have access to $\mathbf{v}_{\text{LT}}$, optimizing a policy that met the $80°$ terminal glideslope constraint required giving the value function an observation containing $\mathbf{v}_{\text{LT}}$. Note that since the value function is not deployed as part of the GN&C system, we can give the value function access to any ground truth state variable. We assume that at the start of the landing segment, the seeker platform is stabilized so that the seeker platform's $w$ axis is perpendicular to the Lunar equipotential surface at the DLS, and $\theta_u$ and $\theta_v$ are both zero. This allows measurement of altitude $h$ and altitude rate of change $\dot{h}$ using the seeker. The ground truth observation given to the value function is shown in Eq. (26) and the observation given to the policy is shown in Eq. (27).

$$\mathbf{o} = \begin{bmatrix} \mathbf{r}_{LT}[2] & \mathbf{v}_{LT} & \mathbf{q} & \omega \end{bmatrix} \tag{26}$$

$$\mathbf{o} = \begin{bmatrix} h & \dot{h} & \mathbf{q} & \omega \end{bmatrix} \tag{27}$$

### D. Network Architecture

The policy and value functions are implemented using four layer neural networks with tanh activations on each hidden layer. Layer 2 for the policy and value function is a recurrent layer implemented using gated recurrent units [5]. The network architectures are as shown in Table 3, where $n_{\text{hi}}$ is the number of units in layer $i$, obs_dim is the observation dimension, and act_dim is the action dimension.

**Table 3  Policy and Value Function network architecture**

|  | Policy Network | | Value Network | |
|---|---|---|---|---|
| Layer | # units | activation | # units | activation |
| hidden 1 | $10 * \text{obs\_dim}$ | tanh | $10 * \text{obs\_dim}$ | tanh |
| hidden 2 | $\sqrt{n_{\text{h1}} * n_{\text{h3}}}$ | tanh | $\sqrt{n_{\text{h1}} * n_{\text{h3}}}$ | tanh |
| hidden 3 | $10 * \text{act\_dim}$ | tanh | 5 | tanh |
| output | act_dim | linear | 1 | linear |

### E. Guidance Segment Optimization

Optimization uses the initial conditions and vehicle parameters given in Section II.A. Fig. 4 plots the reward history (sum of shaping and terminal rewards), with the mean ("Mean R"), mean minus 1 standard deviation ("SD R"), and minimum ("Min R") rewards plotted on the primary y-axis and the mean and maximum number of steps per episode plotted on the secondary y-axis. Similarly, Figs. 5 and 6 plot terminal miss and terminal speed statistics. These statistics are computed over a batch of rollouts (60 episodes). We ran the optimization for 60,000 episodes.

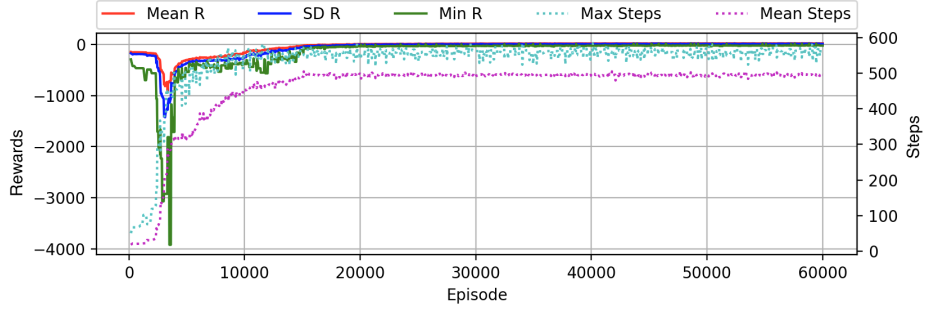

**Fig. 4   Guidance Segment Optimization Reward History**

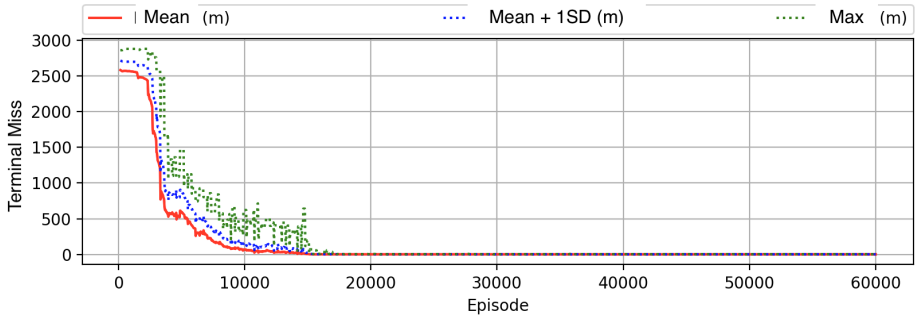

**Fig. 5   Guidance Segment Optimization Miss Distance History**

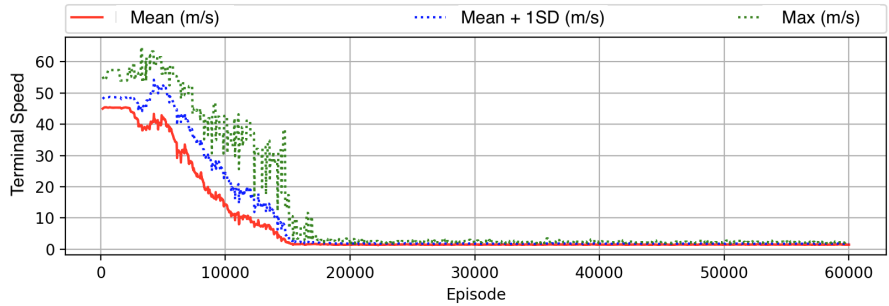

**Fig. 6   Guidance Segment Optimization Terminal Speed History**

### F. Landing Segment Optimization

To generate initial conditions for the landing segment, we ran the optimized guidance policy for 5000 episodes. At the start of each episode in the landing segment optimization, we randomly select a terminal condition from these 5000 episodes to use an initial condition for the landing segment. The landing segment optimization used 120 episode rollouts, and optimization ran for 300,000 episodes. The optimization reward history is plotted in Fig. 7.

## V. Experiments

Once the Meta-RL policy was optimized, we conducted multiple experiments to determine the performance of the Meta-RL integrated G&C system and analyze its generalization capability. For the reader's convenience the labels and descriptions of the test scenarios we carried out are given in Table 4. We measure performance by the success rate, with a successful landing having terminal speed less than 2 m/s with a miss distance less than 10 m, terminal glideslope of at least 80°, the absolute value of pitch and roll less than 10°, and all components of rotational velocity less than 10° / s. The experiment described in the first row of Table 4 has the same conditions as optimization, whereas the following rows test the ability of the policy to generalize to conditions not experienced during optimization.

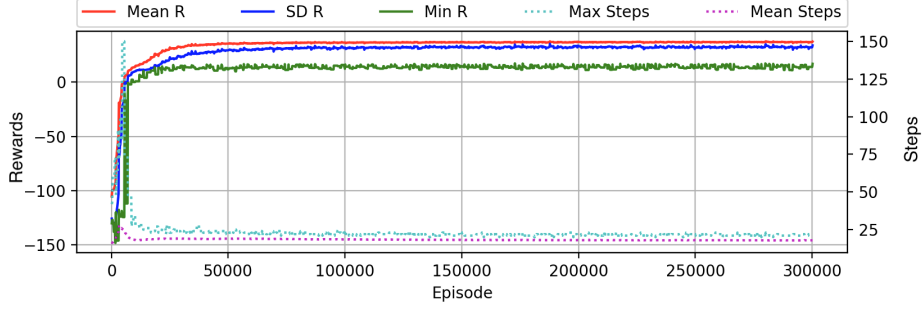

**Fig. 7 Landing Segment Optimization Reward History**

**Table 4 Experiment Descriptions**

| Label | Description |
|---|---|
| Optim | The same conditions used for optimization. |
| AF=$\Delta$ | Branches off of the "Optim" experiment, but at the start of each episode, with probability 0.5, a random engine is chosen for actuator degradation, and that engine's thrust is scaled by $\mathcal{U}(\Delta, 1, 1)$ for the remainder of the episode |
| MV=$\Delta$ | Branches off of the "Optim" experiment, but with the initial vehicle mass set to $1950(1 + \epsilon)$ kg, where $\epsilon$ is randomly drawn from the uniform distribution within the bounds +/- $\Delta$ |
| $\Delta \mathbf{J}_{\text{diag}} = \Delta$ | Branches off of the "Optim" experiment, but the diagonal elements in the lander's inertia tensor are modified by $\Delta$ and the off-diagonal elements by $\Delta/10$, as described in Section II.B |

### A. Performance of Meta-RL Guidance System

To test the meta-RL guidance system, we ran 5000 episodes over the scenarios given in Table 4. In an episode, the G&C switches from the guidance segment policy to the landing segment policy when the altitude referenced to the DLS falls below 5 m. Table 5 tabulates performance over the cases given in Table 4. The max($|\mathbf{q}|$) columns give statistics for either terminal roll or pitch, whichever had the highest $\mu + \sigma$ for that experiment. Similarly, the max($|\boldsymbol{\omega}|$) columns give statistics for the terminal rotational velocity component that had the highest $\mu + \sigma$ for that experiment. $\theta_{\text{gs}}$ is the terminal glideslope.

Figure 8 plots a sample trajectory for the "Optim" case. The first three diverts are clearly visible due to the discontinuity in relative position; the fourth divert is too small to show up clearly in the relative position subplot, but can be seen in the stabilized attitude subplot. The stabilized attitude subplot illustrates the attitude of the stabilized seeker platform, which resets every 2 seconds to align the platform $w$ axis with the line of sight to target (See Section IV.B). The tracking error subplot plots $\mathbf{v}_{\text{err}}$, with the large discontinuity at the end occuring during the landing segment, which uses a policy that does not track $\mathbf{v}_{\text{ref}}$ (the seeker model is not updated during the landing segment). The LOS angles subplot gives the angle in $\mathbb{R}^3$ between various direction vectors, with $\theta_{\text{CV}}$ plotting the angle between the sensor boresight axis rotated into the inertial frame and $\mathbf{v}_{\text{L}}$, $\theta_{\text{RV}}$ plotting the angle between $\mathbf{r}_{\text{TL}}$ and $\mathbf{v}_{\text{L}}$, and $\theta_{\text{CR}}$ plotting the angle between the sensor boresight axis rotated into the inertial frame and $\mathbf{r}_{\text{TL}}$. We see that our reference velocity field formulation does indeed keep $\mathbf{v}_{\text{L}}$ aligned with $\mathbf{r}_{\text{TL}}$. The cost of this formulation is a slight increase in control activity that occurs every 2 seconds when the seeker platform resets. The thrust subplot gives the lander thrust vector in the inertial frame, i.e., $\mathbf{F}_{\text{N}}$ in Eq. (5) from Section II.B. The remainder of the subplots should be self explanatory.

**Table 5   Performance**

| - | Miss (m) | | $\|\mathbf{v}_L\|$ (m/s) | | max($\|\mathbf{q}\|$) (deg) | | max($\|\boldsymbol{\omega}\|$) (deg) | | $\theta_{gs}$ (deg) | | Fuel (kg) | | Success |
|---|---|---|---|---|---|---|---|---|---|---|---|---|---|
| Case | $\mu$ | $\sigma$ | $\mu$ | $\sigma$ | $\mu$ | $\sigma$ | $\mu$ | $\sigma$ | $\mu$ | $\sigma$ | $\mu$ | $\sigma$ | % |
| Optim | 1.1 | 0.39 | 1.54 | 0.13 | 1.3 | 2.6 | 0.6 | 4.5 | 85 | 3 | 186 | 6 | 98.0 |
| AF=0.7 | 1.1 | 0.5 | 1.58 | 0.19 | 1.8 | 3.2 | 0.6 | 4.6 | 84 | 4 | 186 | 6 | 95.1 |
| MV=0.1 | 1.1 | 0.4 | 1.55 | 0.14 | 2.3 | 3.2 | 0.9 | 4.6 | 95 | 3 | 188 | 11 | 97.5 |
| $\Delta \mathbf{J}_{diag} = 30$ | 1.1 | 0.4 | 1.54 | 0.14 | 2.1 | 3.3 | 0.9 | 4.7 | 84 | 4 | 186 | 6 | 96.1 |

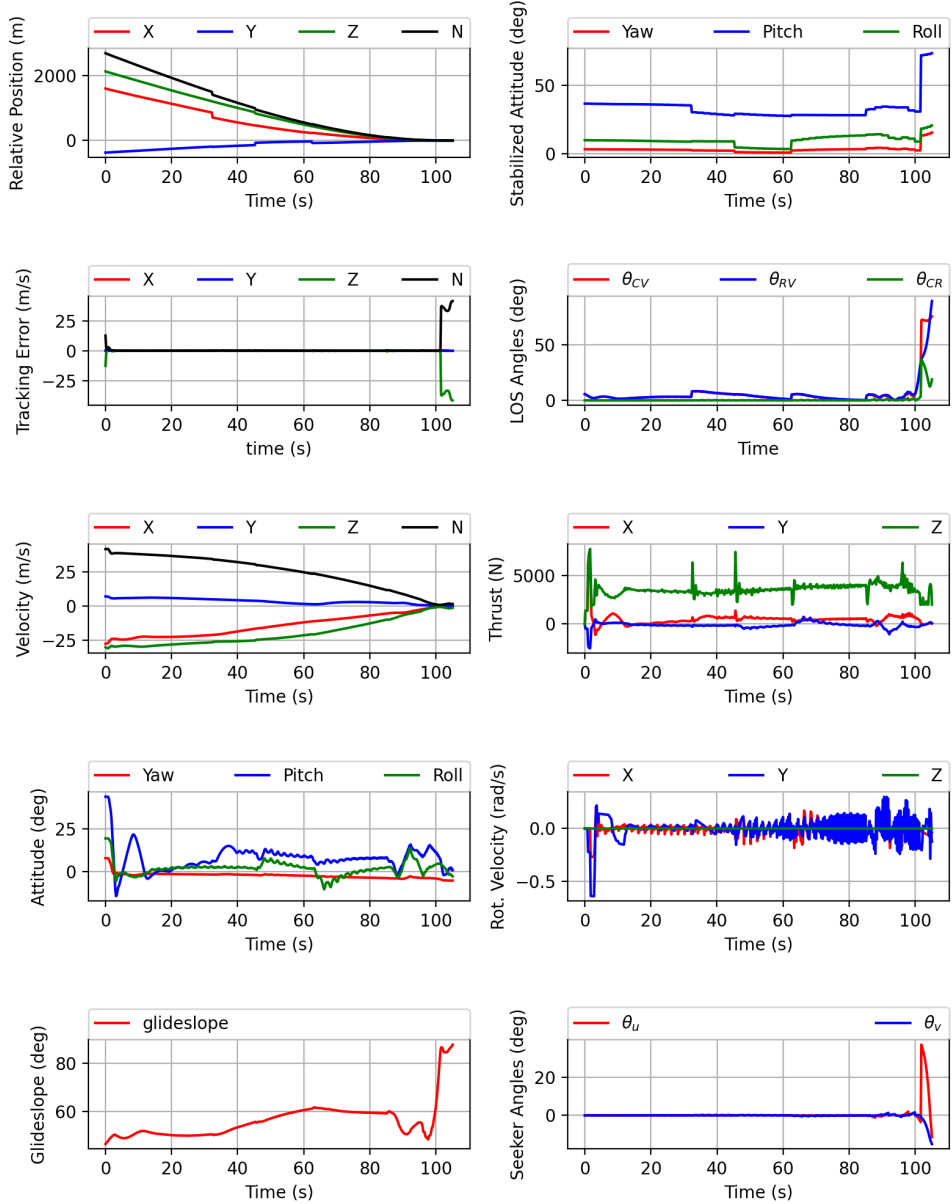

**Fig. 8   Sample Trajectory ("Optim" Case)**

Figure 9 plots a miss distance scatter plot, The downrange bias is because in the guidance segment, the lander is moving along a trajectory that tracks the line of sight to target, but since the end of the guidance segment occurs at an altitude of 5m, the lander is still downrange from the target when the landing segment begins. And since the landing segment attempts to bring the lander vertically to the surface, the lander will end up slightly downrange of the DLS. Figure 10 plots histograms for the terminal speed, glideslope, rotational velocity, and pitch and roll. The "Max" operator indicates for each episode we use the component of the terminal vector with the largest absolute value.

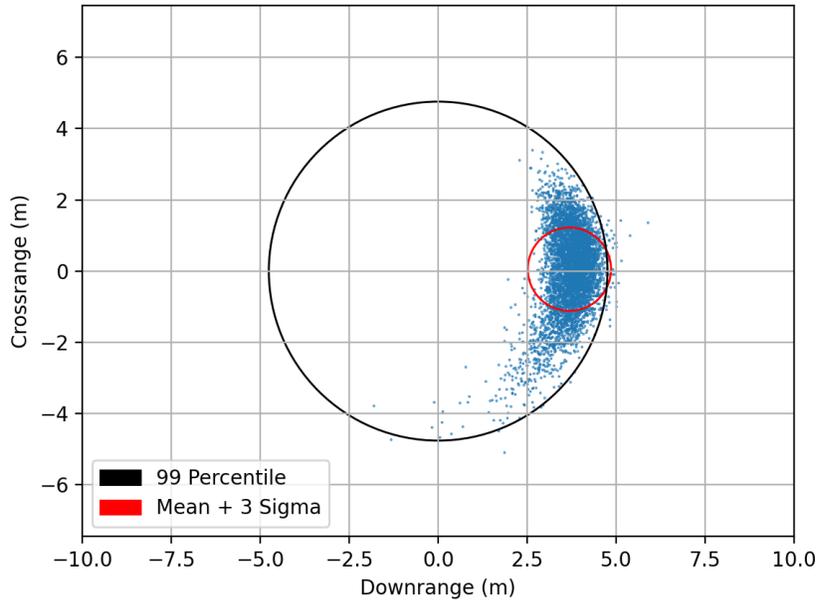

**Fig. 9   Miss Scatter Plot ("Optim" Case)**

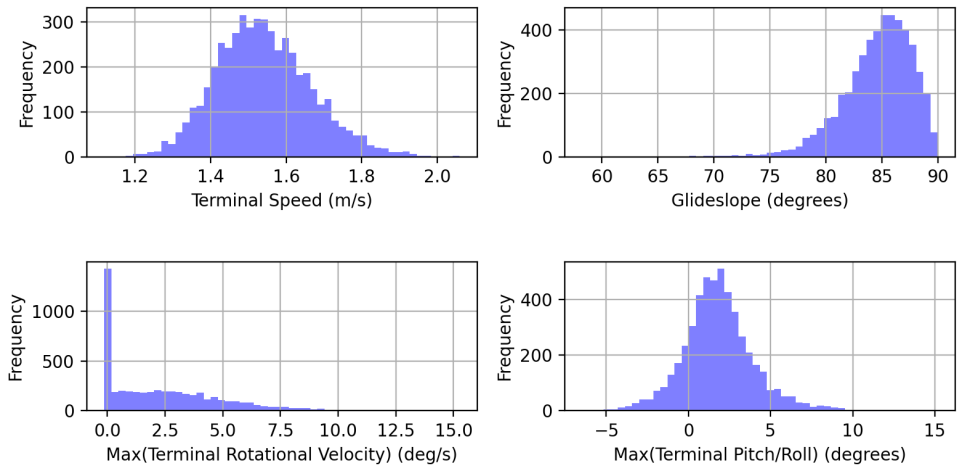

**Fig. 10   Terminal Histograms ("Optim" Case)**

## B. Discussion

When tested under the same conditions used for optimization, the integrated G&C policy met the requirements for a safe landing 98% of the time. From the histograms in Fig. 10, we see that where the system failed to meet the safe landing requirements, the failure cases are not too extreme. It is likely that performance can be improved by refining the coefficients used in the reward function for the landing policy (Eqs. (25a) through (25d)). The policy generalized well to novel conditions not seen during optimization, including actuator degradation, and increased variation in both the lander's initial mass and inertia tensor. It is worth noting that we chose a particularly difficult thruster configuration, in that translational control is coupled with attitude control (there are no dedicated attitude control thrusters).

## VI. Conclusion

We described a guidance, navigation, and control architecture where landing site selection software identifies a safe landing site from images generated by a stabilized seeker. The seeker adjusts its gimbal angles to keep the designated landing site centered in the sensor field of view. We then showed how to formulate a reference velocity field from the seeker gimbal angles, along with closing speed and range to the designated landing site. An integrated guidance and control system was then optimized using meta reinforcement learning, with the guidance and control policy mapping the velocity field tracking error, closing speed, range, rotational velocity, and attitude to commanded thrust for the lander's four thrusters. We then demonstrated that the optimized policy can execute multiple divert maneuvers during the powered descent phase as the landing site selection software refines the designated landing site as sensor resolution increases with decreasing altitude, while meeting the landing requirements ( a level attitude, negligible rotational velocity, and glideslope close to 90°) in 98% of the 5000 test episodes. Importantly, the 2% of the episodes that did not meet the landing requirements, the guidance and control policy came close to meeting the requirements. Further, the optimized policy generalized well to novel conditions not experienced during optimization. To our knowledge, this is the first published work applying missile seeker hardware to the powered descent phase of a planetary landing. Future work will develop the landing site selection software and demonstrate the full guidance, navigation, and control system.

## References

[1] Klumpp, A. R., "A manually retargeted automatic landing system for the lunar module (LM)," *Journal of Spacecraft and Rockets*, Vol. 40, No. 6, 2003, pp. 973–982.

[2] Li, S., Jiang, X., and Tao, T., "Guidance summary and assessment of the Chang'e-3 powered descent and landing," *Journal of Spacecraft and Rockets*, Vol. 53, No. 2, 2016, pp. 258–277.

[3] Roback, V. E., Amzajerdian, F., Bulyshev, A. E., Brewster, P. F., and Barnes, B. W., "3D flash lidar performance in flight testing on the morpheus autonomous, rocket-propelled lander to a lunar-like hazard field," *Laser Radar Technology and Applications XXI*, Vol. 9832, International Society for Optics and Photonics, 2016, p. 983209.

[4] Schulman, J., Wolski, F., Dhariwal, P., Radford, A., and Klimov, O., "Proximal policy optimization algorithms," *arXiv preprint arXiv:1707.06347*, 2017.

[5] Chung, J., Gulcehre, C., Cho, K., and Bengio, Y., "Gated feedback recurrent neural networks," *International Conference on Machine Learning*, 2015, pp. 2067–2075.

[6] Gaudet, B., Linares, R., and Furfaro, R., "Deep Reinforcement Learning for Six Degree-of-Freedom Planetary Powered Descent and Landing," *arXiv preprint arXiv:1810.08719*, 2018.

[7] Scorsoglio, A., Furfaro, R., Linares, R., and Gaudet, B., "Image-based Deep Reinforcement Learning for Autonomous Lunar Landing," *AIAA Scitech 2020 Forum*, 2020, p. 1910.

[8] Gaudet, B., Furfaro, R., Linares, R., and Scorsoglio, A., "Reinforcement Metalearning for Interception of Maneuvering Exoatmospheric Targets with Parasitic Attitude Loop," *Journal of Spacecraft and Rockets*, 2020, pp. 1–14. https://doi.org/10.2514/1.A34841.

[9] Gaudet, B., Linares, R., and Furfaro, R., "Terminal adaptive guidance via reinforcement meta-learning: Applications to autonomous asteroid close-proximity operations," *Acta Astronautica*, 2020. https://doi.org/10.1016/j.actaastro.2020.02.036.

[10] Gaudet, B., Linares, R., and Furfaro, R., "Six degree-of-freedom body-fixed hovering over unmapped asteroids via LIDAR altimetry and reinforcement meta-learning," *Acta Astronautica*, 2020. https://doi.org/10.1016/j.actaastro.2020.03.026.

[11] Siouris, G. M., "Missile guidance and control systems," Springer Science & Business Media, 2004, pp. 102–104. https://doi.org/10.1007/b97614

[12] Finn, C., Abbeel, P., and Levine, S., "Model-Agnostic Meta-Learning for Fast Adaptation of Deep Networks," *ICML*, 2017.

[13] Mishra, N., Rohaninejad, M., Chen, X., and Abbeel, P., "A Simple Neural Attentive Meta-Learner," *International Conference on Learning Representations*, 2018.

[14] Frans, K., Ho, J., Chen, X., Abbeel, P., and Schulman, J., "META LEARNING SHARED HIERARCHIES," *International Conference on Learning Representations*, 2018.

[15] Wang, J. X., Kurth-Nelson, Z., Tirumala, D., Soyer, H., Leibo, J. Z., Munos, R., Blundell, C., Kumaran, D., and Botvinick, M., "Learning to reinforcement learn," *arXiv preprint arXiv:1611.05763*, 2016.

[16] Schulman, J., Levine, S., Abbeel, P., Jordan, M., and Moritz, P., "Trust region policy optimization," *International Conference on Machine Learning*, 2015, pp. 1889–1897.

[17] Kullback, S., and Leibler, R. A., "On information and sufficiency," *The annals of mathematical statistics*, Vol. 22, No. 1, 1951, pp. 79–86.

[18] Shneydor, N. A., "Missile guidance and pursuit: kinematics, dynamics and control," Elsevier, 1998, pp. 101–124. https://doi.org/10.1533/9781782420590

[19] Shneydor, N. A., "Missile guidance and pursuit: kinematics, dynamics and control," Elsevier, 1998, pp. 77,78. https://doi.org/10.1533/9781782420590.

[20] Wan, E. A., and Van Der Merwe, R., "The unscented Kalman filter for nonlinear estimation," *Proceedings of the IEEE 2000 Adaptive Systems for Signal Processing, Communications, and Control Symposium (Cat. No. 00EX373)*, Ieee, 2000, pp. 153–158.